# *Switching friction at a manganite surface using electric fields*

H. Schmidt,[1] J.-O. Krisponeit,[2,3] N. Weber,[1] K. Samwer,[2]
and C. A. Volkert[1,a)]

[1] *Institute of Materials Physics, Georg August University, 37077 Göttingen, Germany*

[2] *1st Physics Institute, Georg August University, 37077 Göttingen, Germany*

[3] *Institute of Solid State Physics, University of Bremen, 28359 Bremen, Germany*

We report active control of the friction force at the contact between a nanoscale asperity and a $La_{0.55}Ca_{0.45}MnO_3$ (LCMO) thin film using electric fields. We use friction force microscopy under ultrahigh vacuum conditions to measure the friction force as we change the film resistive state by electric field-induced resistive switching. Friction forces are high in the insulating state and clearly change to lower values when the probed local region is switched to the conducting state. Upon switching back to an insulating state, the friction forces increase again. Thus, we demonstrate active control of friction without having to change the contact temperature or pressure. By comparing with measurements of friction at the metal-to-insulator transition and with the effect of applied voltage on adhesion, we rule out electronic excitations, electrostatic forces and changes in contact area as the reasons for the effect of resistive switching on friction. Instead, we argue that friction is limited by phonon relaxation times which are strongly coupled to the electronic degrees of freedom through distortions of the $MnO_6$ octahedra. The concept of controlling friction forces by electric fields should be applicable to any materials where the field produces strong changes in phonon lifetimes.

Friction is a complex phenomenon that occurs between two bodies at a sliding contact. Despite the fact that it often can be described by straight-forward empirical relations, its fundamental cause is by no means simple. With the advent of the atomic force microscope (AFM), understanding and controlling nanoscale friction has become one of the major interests in modern tribology. A promising direction is reported in several literature studies [1-8] which show a clear change in measured nanoscale friction force when the electronic state of the material is altered. Abrupt increases are observed in the non-contact dissipation of Nb [5] and the contact friction of YBCO [6] and Pb [1,8] as the materials are heated through their superconducting transitions. Contact friction measurements on Si [4] and GaAs [2] semiconductors demonstrate a strong dependence on the charge carrier density, while the contact friction of $VO_2$ is strongly increased on heating through the metal-to-insulator transition (MIT) from the insulating to the metallic state [3,7].

---

[a] Author to whom correspondence should be addressed. Electronic mail: volkert@ump.gwdg.de

Although the literature consistently reveals an impact of the electronic state on friction, a clear understanding of the origin of the effect has not yet emerged. The available theories based on the conversion of mechanical energy into electronic excitations predict only small increases in friction when the number of accessible electronic states are increased, while the experiments often reveal much larger changes in friction. As a result, most studies have explained the effect of changes in the electronic state on friction through electrostatic interactions. Coulomb and capacitive forces at the surfaces of dielectrics (including at surface oxides on conductors) can easily be large enough to explain the observed changes in friction [2-4,6,9]. But even in the absence of a full explanation, the pervasiveness of the effect of electronic states on friction raises the intriguing question of whether they can be used to control friction.

Most of the literature studies up until now have used temperature [5,6], contact pressure [3,7], or a sustained bias voltage [2,4,10] between the tip and the sample to control the electronic state of the material. From a practical point of view, it would be advantageous to find a way to control friction without having to change a process parameter such as pressure or temperature or without having to maintain a voltage between the two contact surfaces.

In this study, we report on active control of friction by electric field-induced resistive switching of the surface region of a manganite film. Perovskite manganites offer a range of different electronic state transformations [11], including the well-known temperature driven metal-to-insulator transition [12], the colossal magnetoresistance (CMR) effect [13], as well as bipolar nanoscale resistive switching driven by electric fields [14]. Here we use the last to reversibly alter the resistive state of a nanoscale region of a perovskite manganite surface to investigate the effect on friction using AFM-based methods. In order to get at the origins of the observed changes in friction due to resistive switching, we compare it with the friction behavior observed on heating the film through the metal-to-insulator transition as well as with the effect of applied voltage on the adhesive force.

A manganite $La_{0.55}Ca_{0.45}MnO_3$ thin film was deposited on a MgO substrate with (100)-orientation by the metalorganic-aerosol deposition technique [15]. Small angle x-ray scattering gives a film thickness of approximately 44 nm and θ/2θ x-ray diffraction experiments confirm (100) epitaxial growth on the substrate and a lattice constant of 3.87(3) Å (see Fig. S1). The film sheet resistance vs. temperature behavior probed using four-point measurements shows a clear metal-to-insulator transition at $T_{MI} = 245$ K (see Fig. S2). The measurements are consistent with the phase diagram for a Ca doping fraction of 0.45 which shows a transition from a ferromagnetic metal to a paramagnetic insulator at 245 K [12]. The material is well known to show resistive switching, with a polarity and behavior depending on the electrode materials [16]; it also can be switched on the nanoscale by conductive atomic force microscopy (C-AFM) [17,18].

Friction force measurements and resistive switching have been performed in an ultra-high vacuum chamber with an Omicron VT-AFM/STM at a base pressure of $10^{-10}$ mbar. For the friction experiments we performed friction force microscopy (FFM) using standard platinum coated Si-cantilevers (MikroMasch DPE17/Al BS) with normal and torsional spring constants of 0.2 N/m and 23.49 N/m, respectively. The normal force was kept constant at $F_N = 5$ nN and the scan velocity was 2000 nm/s. The lateral forces were calibrated using analytical procedures described in Ref. [19]. The combination of FFM and C-AFM allows us to simultaneously measure the friction force and to switch the resistive state of the manganite back and forth between the metallic (low resistance state LRS) and insulating states (initial state IS, high resistance state HRS) by applying a voltage larger than the switching voltage $|V_C| \approx 3$ V [17]. Additionally, C-AFM gives us the opportunity to resolve the resistive state by recording current maps at lower voltages during friction measurements. In order to avoid plastic protrusions that can be generated by Joule heating under application of the large positive switching voltages used for generating the LRS [17], we rapidly scanned the area of interest several times under an applied bias rather than using prolonged voltage pulses at a fixed cantilever position. Using this method, we avoided any noticeable changes in the surface morphology in our experiments. Nor was a correlation between RMS surface roughness (typically around 1 nm) and friction or current observed.

In each resistive switching experiment, the friction forces were measured in a 500x500 $nm^2$ region on the sample surface by performing 600 friction loops, i.e. by recording the lateral forces experienced by the cantilever tip while scanning forward and backward traces over the sample surface. At the same time, the surface topography was recorded from the vertical deflection of the cantilever. The materials resistive state was also probed during the friction loops by recording the current from an applied voltage of 0.1 V which is well below the threshold needed for switching. First, friction forces, topography, and current maps were obtained from a 500x500 $nm^2$ region of the specimen in the insulating initial state (IS). Then the near surface region was switched to the conducting LRS by rapidly and repetitively scanning over the same region with a tip voltage of +3.5 V [17,18]. After obtaining friction forces, topography, and current maps from the switched state, the region was switched back to an insulating HRS using a tip voltage of -3.5 V, and friction, topography, and current maps were obtained once again using a voltage of 0.1 V, which is well below the switching threshold. This experiment was repeated for several different local regions of interest on the specimen. The current maps confirmed a mostly uniform change in resistive state in the switched region (Fig. S3). However, because the current amplifier reached its saturation limit of 50 nA even at a low bias (0.1 V) in the LRS and because the currents in the insulating states (IS and HRS) are below the detection limit of 0.1 nA, we obtained only threshold information about the resistive state.

Fig. 1(a-c) shows topography and lateral force maps of the three states (IS, LRS, and HRS), for a representative region on the sample. Due to some drift between the tip and specimen during the resistive switching procedure, the scanned regions are slightly displaced relative to each other and have been aligned using the topography maps. The common height and force scales for the three sets of maps show clearly that there are no significant changes in topography, but that the lateral forces are strongly reduced in the conducting LRS (Fig. 1(b)). Variations in the lateral forces are observed within each map that are on a similar length scale as the topography (Fig. 1(a-c)).

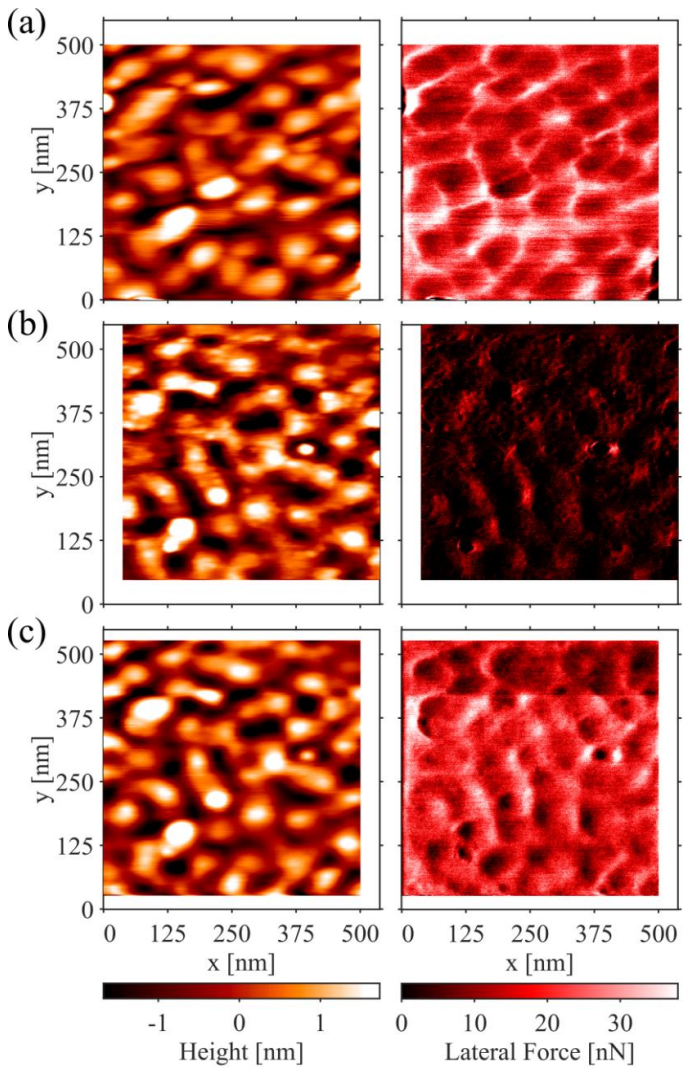

**FIGURE 1:** AFM-based measurements of $La_{0.55}Ca_{0.45}MnO_3$. (a-c) Topography (left) and lateral force (right) maps obtained from forward traces measured on (a) the initial insulating state (IS), (b) after resistively switching to the metallic state (LRS), and (c) after resistively switching back to the insulating state (HRS).

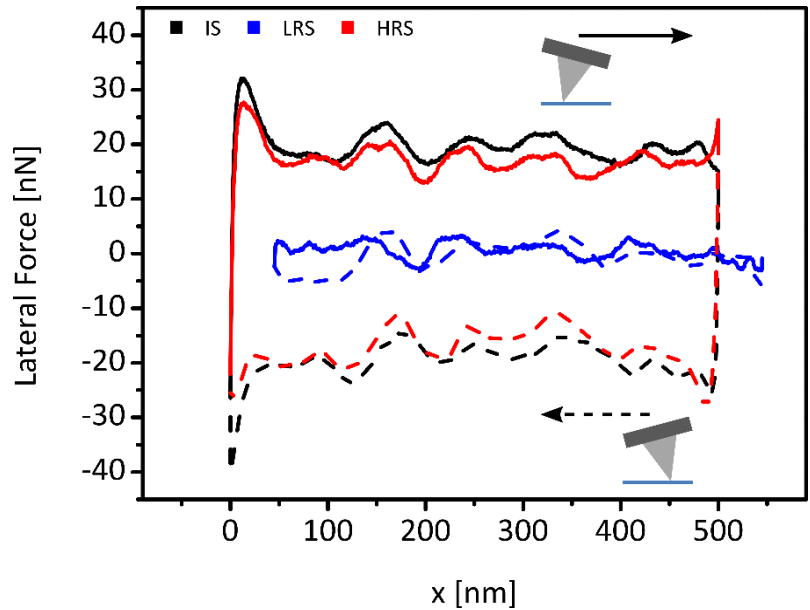

**FIGURE 2:** Average friction loops for the insulating states (IS = black; HRS = red) and the conducting state (LRS = blue). The friction is represented by the difference between the forward (solid line) and backward (dashed line) trace lateral forces divided by two. Thus, the friction is high for the insulating states and low for the conducting state. The slight offset (ca. 15 nm) in the x-direction between the forward and backward scans is due to rotation of the tip on reversing the scan direction.

The dramatic change in friction forces with resistive state is illustrated most clearly by the friction loops. Fig. 2 shows friction loops for each resistive state, which have been calculated by averaging the lateral forces for all 600 forward and backward traces within a single region of interest. The exact position of the AFM tip deviates between forward and backward scan due to torsion of the cantilever torsion and is corrected by determining the maximum correlation between trace and retrace of the topography maps [20]. The x-shift of around 50 nm of the LRS loop relative to the IS and HRS has a different origin; it is due to displacement drift that occurred during resistive switching. A clear hysteresis in the lateral force and thus large friction is observed for the initial and switched insulating states (IS and HRS), while the conducting state (LRS) shows almost no hysteresis and thus low friction.

Large variations in lateral force are observed along the forward and backward traces that are correlated with sample position (Fig. 2). These local variations are mostly unchanged on switching to the conducting state and on switching back to the insulating state, and thus must be caused by spatial heterogeneities in the specimen. The most likely explanation comes from the gradients in the local sample surface height dh/dx which contribute additively to the measured lateral force [21].

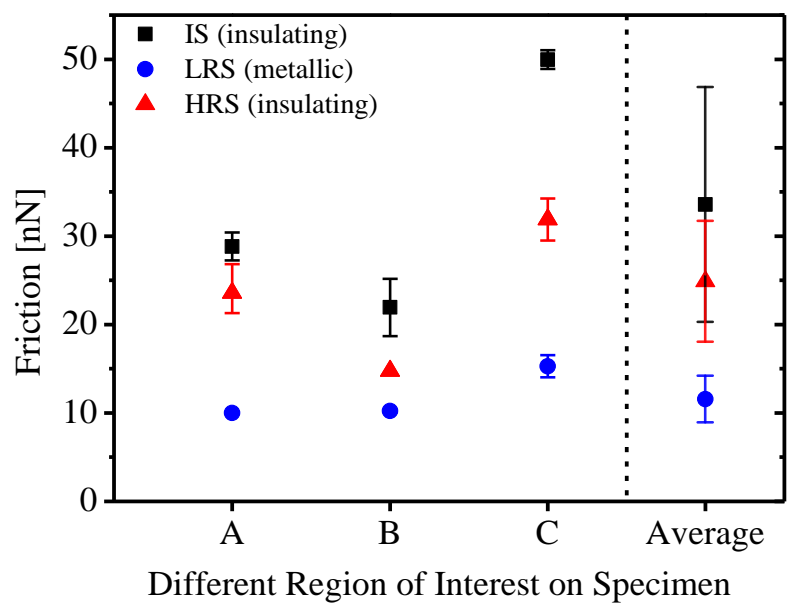

FIGURE 3: (color online) Friction of the insulating state is consistently higher than of the metallic state in three different regions A, B, and C of the $La_{0.55}Ca_{0.45}MnO_3$ film. The left hand portion of the plot shows average values (symbols) and ranges (bars) for friction of each resistive state in the three regions. The means and standard deviation of the position averaged values of the friction for each resistive state are shown in the right hand portion of the plot.

In order to confirm the robustness of the effect of the resistive state on friction forces, the above experiment was repeated for three different regions of interest on the specimen (Fig. 3). The average friction force for each 500x500 $nm^2$ region of interest and resistive state was calculated by shifting and subtracting the forward and backward lateral forces and dividing by two. Despite a clear dependence of the initial friction on location, the friction for all regions is clearly reduced on switching from the insulating IS to the conducting LRS and increases on switching back to the HRS. However, it is found that the friction force of the IS is not fully recovered on switching back from the LRS to the HRS. This observation is consistent with previous studies [17] where resistance fatigue was observed when switching a manganite thin film back and forth between its resistive states, particularly after the first switching cycle. Specifically, the difference between the resistances of the two states decreased with switching cycle number. Furthermore, Krisponeit et al. reported local variations in the threshold voltage and the magnitude of the resistance change on switching [17] as well as in the time stability of the LRS [18]. It is possible that the variations from region to region in the initial values of the friction and in the magnitude of the friction force change on switching may have the same origins as the local resistance variations. Whether the causes of resistance and friction variations are local variations in composition, surface charge density, thickness of the oxygen deficient surface or dead layer [23,24], or some other effect, is not yet clear.

In order to further investigate correlations between the near-surface properties and friction of the LCMO film, the friction force near the metal-to-insulator transition at 245 K was measured (Fig. 4). FFM maps were performed with the same methodology as above from an IS sample for temperatures between 110 and 300 K using both silicon and Pt-coated silicon cantilevers. Measurements were performed at room temperature at the beginning and at the end of the temperature series to confirm that there had been no irreversible changes in the specimen or the AFM tip. The friction force measured with the Si tip decreases continuously with increasing temperature while the friction force measured with the Pt-coated tip remains constant within the noise. The lateral forces were not calibrated for the tip used in these measurements so that the friction forces are reported in arbitrary units. Neither measurement series shows any effect of the metal-to-insulator transition on the friction.

Changing the electronic properties by resistive switching causes a clear change in friction, while changing the electronic properties by a temperature-driven metal-to-insulator transition has no effect. The likely explanation for this is the approximately 1 nm thick high resistivity surface dead layer [23,24]. It is often argued that the surface dead layer is changed during resistive switching and may undergo a layer-by-layer electronic structural transformation [25] and/or exchange oxygen vacancies with the underlying film by electrochemical migration [14,16]. Both mechanisms are expected to produce changes in the near-surface structure and electronic properties. In contrast, the surface dead layer remains intact through the metal-to-insulator transition and is sufficiently thick to shield any contributions from the underlying film material. This temperature–dependent behavior is in excellent agreement with the so-called thermolubric effect [26,8], which attributes the widely observed thermally-activated friction at cryogenic temperatures to adhesion hysteresis from making and breaking contacts between hard materials during sliding. A fit to the friction forces measured with a Si-tip give an activation energy of 0.2 eV which is within the range of those reported in the literature [26]. As predicted by the theory, thermolubricity is not observed for the Pt-coated tip since the relatively easy deformation of the Pt interferes with the making and breaking of adhesive contacts [26,8]. Thus, contact dynamics appear to entirely control the temperature dependence of the friction response in this case, while the metal-to-insulator transition in the buried film is not detected. It is conceivable that contributions from the transition in the underlying material might be detected more strongly for larger normal forces [8, 27].

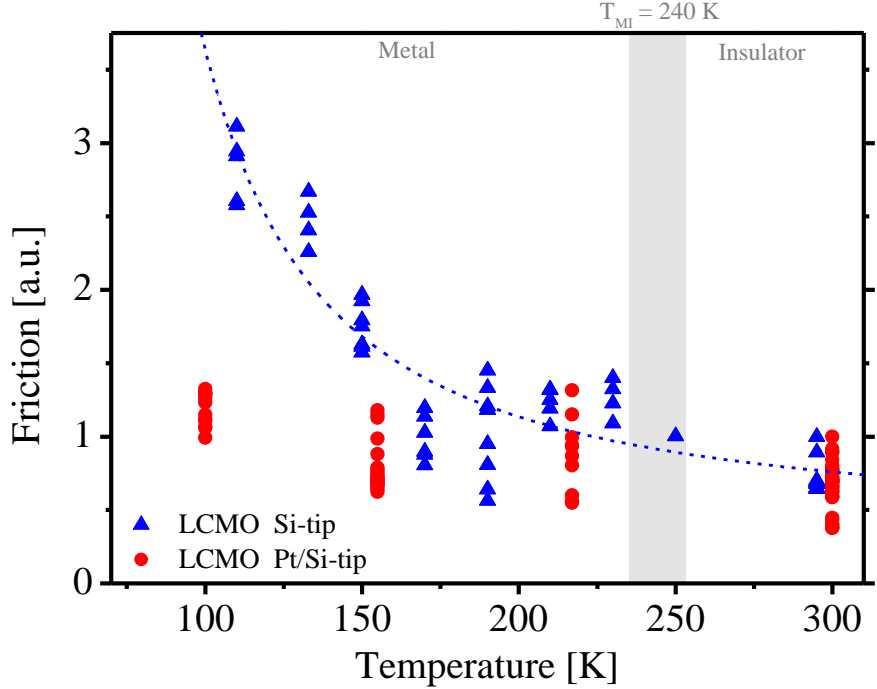

F**IGURE 4:** (color online) Friction force versus temperature for the La$_{0.55}$Ca$_{0.45}$MnO$_3$ film measured with a silicon (blue triangles) and a Pt-coated silicon tip (red circles). The dashed line shows a thermolubricity model fit to the silicon-tip data with an activation energy of 0.2 eV.

Contact sliding of two surfaces against each other causes a variety of different electronic interactions including Ohmic losses and generation of electronic excitations [2,28]. However, theoretical estimates of the magnitudes of these effects yield values no more than $F/v = 10^{-8}$ Ns/m, where F is the excess friction force due to the change in resistive state and v the scanning velocity [2]. Excess friction forces have been reported frequently in the literature and have been attributed to a variety of effects including additional charge carriers [2,4], the superconducting transition [1,5,6,8] (the Cooper pairs in the superconducting state are assumed to be too tightly bound to be excited by friction [5]), and metal-to-insulator transitions [3,7]. However, the magnitude of the electronic effect observed for experimental contact friction measurements is many orders of magnitude larger than predicted by theory [28]. Even if one considers stick-slip motion of the tip, which would lead to intermittent substantially higher sliding speeds [29], theoretical predictions of electronic effects and slip velocities cannot explain the observed excess friction force. Only in cases such as non-contact dissipation measurements [5] and dissipation during adsorbate sliding [1,30], does the measured contribution show qualitative agreement with theory.

The disagreement between theory and contact friction experiments has largely been reconciled in the literature by considering the electrostatic forces that are generated between the two contacting materials as a result of contact electrification or tribo-electrification. These forces are in addition to any applied forces and will change the contact area between the two bodies. Particularly in oxides and other insulating materials, it has often been proposed that electrostatic forces from surface and other trapped charges may overwhelm other effects [2–4,6,9]. For example, Qi et al. [2] were able to obtain quantitative agreement with their experimental excess friction, which was on the order of $10^{-5}$ Ns/m, by accounting for the electrostatic forces from trapped charges in the surface oxide layer. Altfeder et al. [6] found even larger excess dissipation levels in the range of $10^{-2}$–$10^{-3}$ Ns/m between the normal and superconducting phase of YBCO, and attributed it to contact electrification and electrostatic effects in the oxygen depleted surface layer of the normal phase. Kim et al. [3] also employed this argument to explain the large excess friction observed during pressure- and temperature-induced transformation of $VO_2$ from the insulating to metallic phase. Specifically, they attributed the excess friction to trapped charges in the surface $V_2O_5$ dielectric layer on the metallic $VO_2$ domains, leading to higher Coulomb attraction between the tip and sample. Finally, although preferring an explanation based on electron-phonon coupling, a recent study of the friction at the superconducting transition temperature of Pb also cannot rule out contributions from electrostatic forces [8].

In our studies of resistively switched friction, not only the magnitude of the excess friction due to changing the resistivity (F/v = $10^{-3}$ - $10^{-2}$ Ns/m), but even the direction of the friction change contradicts most existing theories for electronic excitations. We observe a clear <u>reduction</u> in friction on switching from the insulating to the conducting state. Thus, it seems highly unlikely that the friction changes observed here have their origins in electronic excitations. Instead, possible contributions from electrostatic forces to LCMO friction were tested by measuring the effect of an external voltage on adhesion for both the IS/HRS and LRS (Fig. 5). Force-distance measurements were performed at fixed locations on the film in both the HRS and LRS to obtain the adhesion (pull-off) force as a function of the voltage applied between the LCMO film and a Pt-coated AFM tip. The applied voltage was kept below the threshold voltage to avoid unintentionally resistively switching the film. Pull-off occurs during tip retraction when the cantilever stiffness first exceeds the gradient in the tip-specimen interaction force. For the relatively compliant cantilever used here, pull-off occurs close to the point of maximum attractive force between the tip and specimen. Any changes in pull-off force with applied voltage are due to electrostatic forces between the Pt-coated Si tip and the specimen and the resultant increase in the contact area between the tip and specimen. Note that the tip used for this study was different from the ones used to measure the lateral friction force during resistive switching (Figs. 1 to 3) so that the magnitude of the forces will generally not be directly comparable.

The adhesive force is found to be similar for the two resistive states and increases slightly with applied voltage (Fig. 5). The data have been fit with a simple model based on a Hertzian contact and parallel plate capacitance between the tip and specimen (see Supporting Information). The material parameters needed to fit the model fall within the range of parameters found in similar materials in the literature. The roughly 50% increase in the adhesion force at the maximum applied voltage results from a 28% increase in the contact area and is split with a ratio of 1:8 between electrostatic forces and non-electrostatic adhesive forces. The fact that the minimum adhesive forces are comparable in the two resistive states indicates that the elastic constants [31] and the effective tip - specimen capacitances are not significantly different. The shift of the voltage at the minimum by around 0.7 V between the two resistive states indicates a change in either the specimen work function and/or a shift in the density of near-surface trapped charges, such as from extrinsic interface charges at the buried interface between the dead layer and the film.

The electrostatic forces are expected to also affect friction by increasing the normal force, which increases the contact area. Ignoring possible lateral electrostatic forces [2] for the moment, accounting for the observed effect of resistive switching on friction would require that the contact area between the tip and the HRS be around three times larger than to the LRS. Under the friction measurement conditions used here, where the specimen is only electrically connected through the tip, the electrostatic forces come from the contact potential difference between the tip and specimen and from any trapped charges or charges generated by tribo-electrification (which may be quite different from those generated by contact electrification [32]). However, since spontaneous resistive switching was not observed during the friction measurements, the maximum electric potential generated during friction must be less than the threshold voltage for switching, leading to a maximum change in contact area of 28%. We therefore believe we can rule out electrostatic forces as the dominant cause of the effect of resistive switching on the LCMO friction.

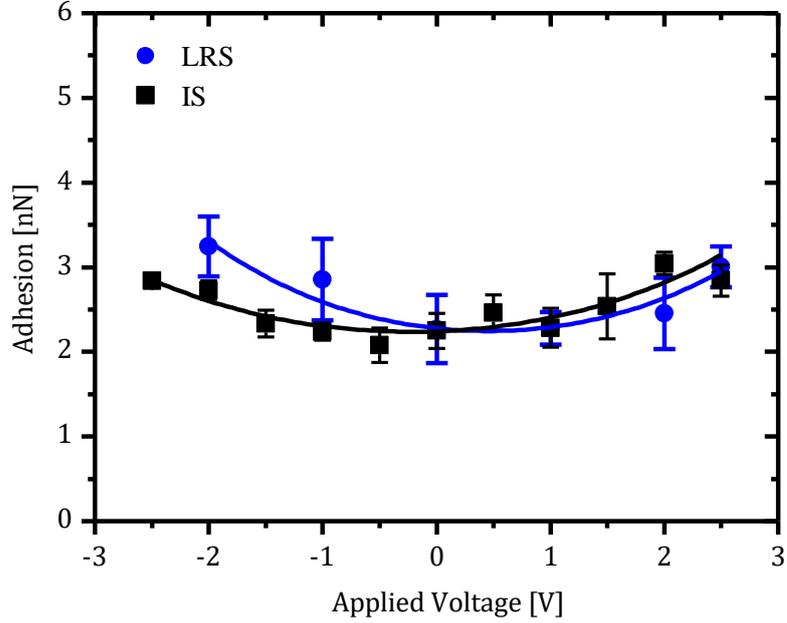

FIGURE 5: (color online) Adhesion between the tip and a $La_{0.55}Ca_{0.45}MnO_3$ film as a function of the applied voltage for the insulating (IS) and conductive states (LRS) as determined from the pull-off force from five individual force-distance curves per voltage with their standard deviation as error. Solid lines are bet fits to the data including effects of capacitive forces and changes in contact area (see Supporting Information)

Having ruled out electronic excitations, electrostatic interactions, and changes in contact area (also due to changes in elastic properties [31]) as possible causes for the large and reversible change in friction due to resistive switching, we are left with dissipation by electron-phonon excitations as a possibility. Most of the systems where friction has been observed to change with the electronic state also involve changes in the phonon behavior. For example, the structural phase transformation in $VO_2$ and the superconducting transition involve changes in the phonon density of states as well as in the electron-phonon coupling. Also for $La_{0.55}Ca_{0.45}MnO_3$, the phonon density of states changes quite strongly at the metal-to-insulator transition due to the formation of Jahn-Teller excitations. A model for the direct coupling of the coherent interface slip pulses that occur in the contact interface during sliding friction to phonons [33,34] shows that the phonon lifetime $\tau_{ph}$ determines the rate at which the energy of the slip pulses is spread and dissipated in the material. This leads to a friction force that is proportional to the phonon relaxation rate $F/v \propto 1/\tau_{ph}$. Phonon relaxation rates can be estimated around the metal-to-insulator transition from thermal conductivity measurements, since the phonon contribution is expected to dominate over electron and magnon contributions. Changes in thermal conductivity and phonon lifetimes with temperature in various manganites are known to scale with static $MnO_6$ distortions [35]. For example, phonon scattering rates are estimated to decrease by a factor of 2.7 between the insulating phase (300K) and the metallic phase (50K) of $La_{0.7}Ca_{0.3}MnO_3$ [35], suggesting that the friction should decrease by the same factor. Assuming the $MnO_6$ distortions are similar in the equilibrium low temperature metallic phase and the metastable resistively switched LRS phase, the phonon lifetime model provides an excellent explanation for both the sign and magnitude of the friction change observed here. Since distortion of the $MnO_6$ octahedra results from the interplay of lattice, charge, and spin degrees of freedom, strongly correlated materials such as the manganites offer promising systems for controlling friction.

In conclusion, we have demonstrated control of nanoscale friction by resistively switching a manganite film using an electric field. A reproducible and reversible change in friction force is observed on changing between high and low resistance states. Contributions from pure electronic excitations, electrostatic interactions, and changes in contact area have been ruled out as the cause of the friction change. However, direct coupling of coherent slip pulse excitations in the interface to phonons in the surrounding material, which have different lifetimes in the metal and insulator phases, does provide a good explanation for the observations. Thus, materials with strong electron-phonon or spin-phonon coupling are promising candidates for controlling friction by external fields [10]. Facile switching of friction forces would offer new control concepts for self-propelled directed motion as well as energy dissipation minimization at contact interfaces.

See supplementary material for a θ-2θ x-ray diffraction spectrum and 4-point resistance measurement versus temperature of the $La_{0.55}Ca_{0.45}MnO_3$ film, for current maps in the different resistive states, and for a model of the effect of voltage on adhesion forces.

This work is part of the CRC 1073 (project A01) funded by the Deutsche Forschungsgemeinschaft (DFG). JOK acknowledges support by the Institutional Strategy of the University of Bremen, funded by the German Excellence Initiative. The authors thank A. Belenchuk for sample preparation and C. Meyer for x-ray diffraction characterization, and gratefully acknowledge helpful discussions with B. Damaschke, J. Krim, and V. Moshnyaga.

## *Supporting Information*

for *Switching friction at a manganite surface using electric fields*

by H. Schmidt, J.-O. Krisponeit, N. Weber, K. Samwer, and C. A. Volkert

**Thin Film Characterization**

θ-2θ x-ray diffraction was used to investigate the crystallographic texture of the 44 nm thick $La_{0.55}Ca_{0.45}MnO_3$ film on the (100) MgO substrate [S1]. Only peaks associated with (100) planes are observed, supporting an epitaxial relation between film and substrate. The metal-to-insulator transition temperature was obtained from 4-point ppms measurements of the film resistivity and is in good agreement with transition temperatures found in the literature for this composition [S2].

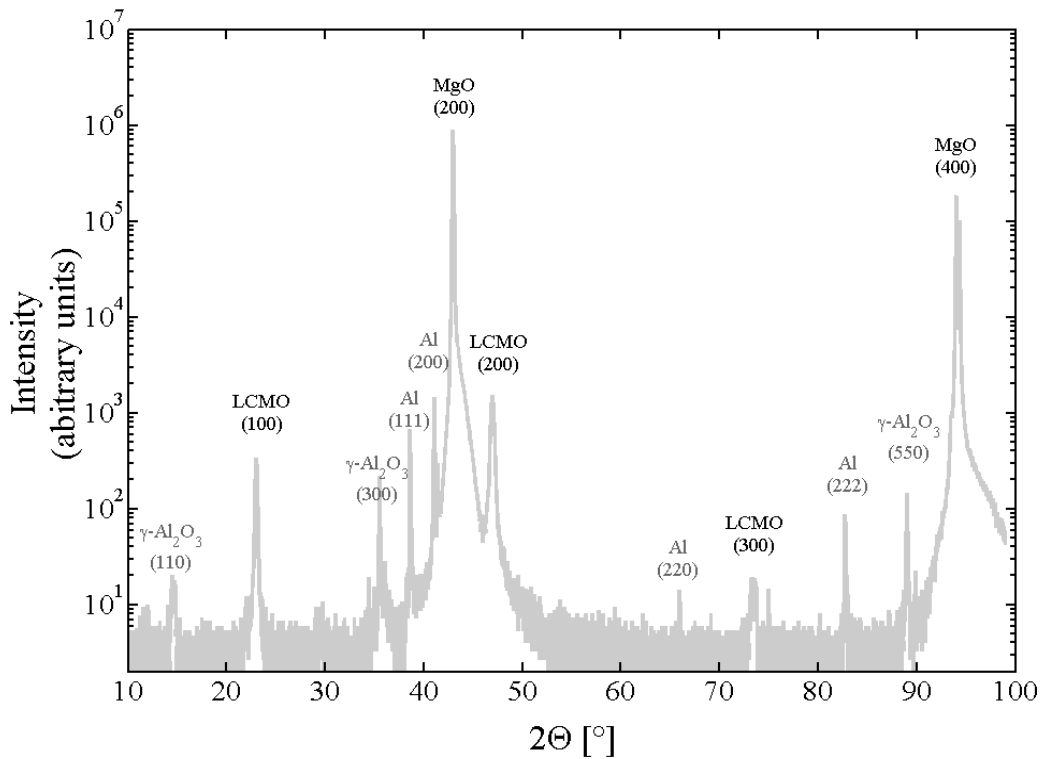

**FIGURE S1:** θ-2θ x-ray diffraction experiments confirm (100) epitaxial growth of the $La_{0.55}Ca_{0.45}MnO_3$ thin film on the MgO substrate. The Al and $Al_2O_3$ peaks come from the sample holder.

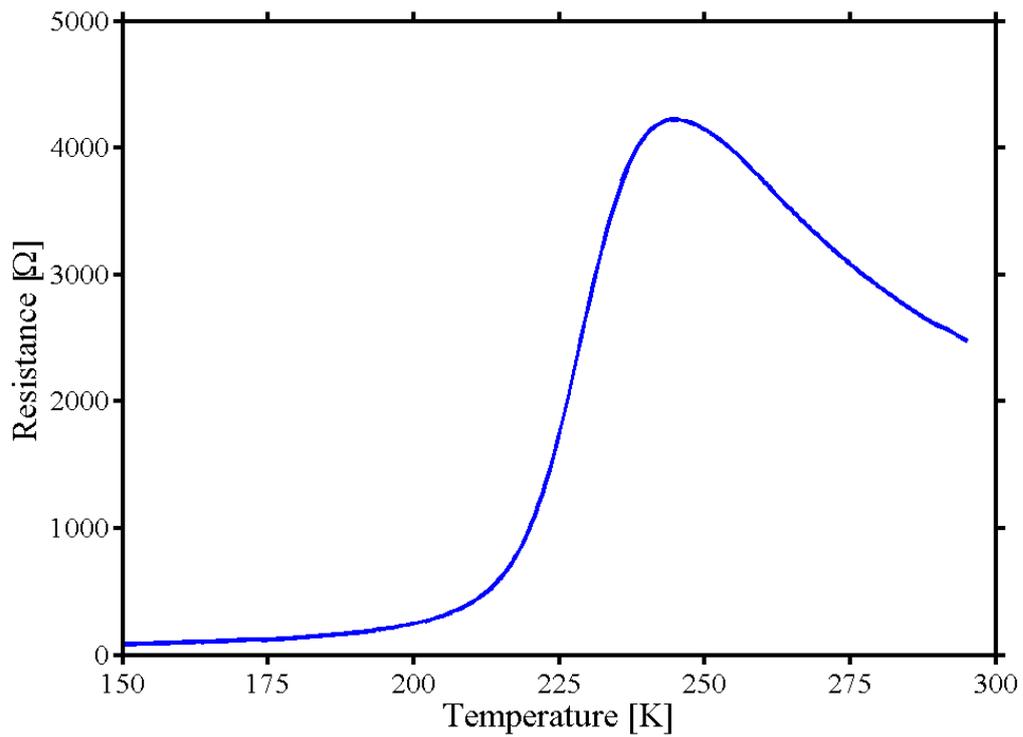

**FIGURE S2:** Four-point measurement of the $La_{0.55}Ca_{0.45}MnO_3$ film resistance as a function of temperature shows a clear metal-to-insulator transition at $T_{MI} = 245$ K.

**C-AFM Measurements**

Current maps were obtained simultaneously with the friction forces and topography in 500x500 nm$^2$ regions on the sample surface by applying a voltage of 0.1 V to the Pt coated tip, which is well below the threshold needed for switching [S3,S4]. Since the current amplifier reached its saturation limit of 50 nA at 0.1 V in the LRS and the currents were below the detection limit of 0.1 nA at 0.1 V in the insulating states (IS and HRS), we only obtained threshold information about the resistive states.

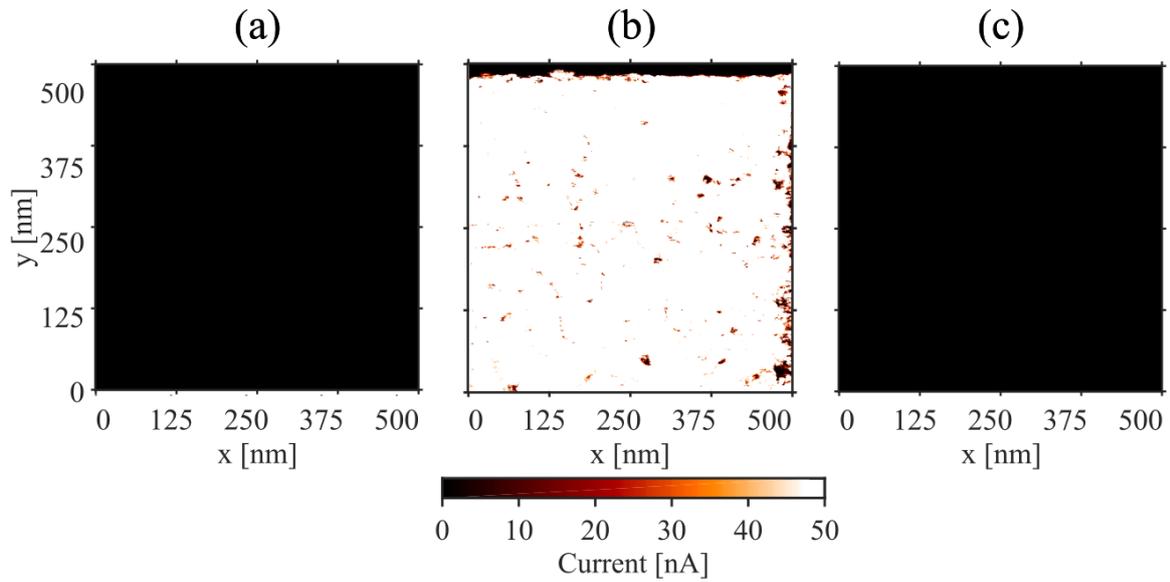

**FIGURE S3:** C-AFM measurements of the (a) IS, (b) LRS, (c) HRS of the La$_{0.55}$Ca$_{0.45}$MnO$_3$ thin film sample. The current maps confirm a mostly uniform change in resistive state in the switched region

**Model for Effect of Capacitive Forces on Adhesion**

The application of a voltage between a contacting conducting tip and dielectric specimen results in electrostatic forces that contribute to the interatomic forces. The additional attractive electrostatic force increases the contact area between the tip and the specimen and the stresses in the specimen under the tip. In the simplest case of an elastic Hertzian contact and parallel plate capacitance, we can write the total force of adhesion between the tip and the specimen as [S5-S7]

$$F_{ad} = A \cdot \left(f_{ad} + \frac{\varepsilon(V-V_o)^2}{2d^2}\right),$$

where $A = \pi \cdot (3F_{ad}R/4E^*)^{2/3}$ is the Hertzian contact area, $f_{ad}$ is the adhesive force per unit area in the absence of any electrostatic effects, $\varepsilon$ is the permittivity of the contact region, $d$ is the thickness of the contact interface, $R$ is the radius of curvature of the tip, and $E^*$ is the reduced modulus of the contact. $V_o$ is the voltage at minimum adhesion force and reflects the contact potential difference between the tip and specimen. Any possible electric field-induced and stress-induced changes in the material properties are ignored, which may also contribute to adhesion by changing the modulus or through changes in the surface processes that determine the adhesion hysteresis.

This equation provides a reasonable fit to the measured adhesion force as a function of voltage (Figure 4). Assuming typical values for the tip radius $R = 50\ nm$ and for the reduced modulus $E^* = 10^{11}\ N/m^2$, we obtain a contact area of $A = 2.7\ nm^2$ and an average contact stress of $f_{ad} = 8.2 \times 10^8\ N/m^2$ at the minimum and a contact area of $A = 3.5\ nm^2$ and an average contact stress of $(f_{ad} + \varepsilon(V-V_o)2/2d^2) = 9.2 \times 10^8\ N/m^2$ at the maximum measured adhesion force. These numbers are in good agreement with the typical values quoted for contact areas and average stresses at AFM tips [S5]. The roughly 50% difference between the minimum and maximum adhesion force results from a 28% increase in the contact area and from a 13% increase in the average contact stress. The change in average contact stress between minimum and maximum adhesion is given by $\varepsilon(V-V_o)2/2d^2 = 1.0 \times 10^8\ N/m^2$. Setting $V - V_o = 1.5\ V$ and $d = 1\ nm$ [S8], we obtain a relative permittivity in the contact interface of 10, which is also quite reasonable [S9]. The shift of the minimum voltage by around $\Delta V_o = 0.7\ V$ between the two resistive states indicates a change in either the specimen work function and/or a shift in the density of near-surface trapped charges. A parallel plate capacitor model would predict that a trapped charge density of $\sigma = 2\varepsilon\Delta V_o/d = 0.12\ C/m^2 = 8 \times 10^{17}\ e/m^2$ at the buried interface of a $d = 1\ nm$ thick surface dead layer could account for the observed shift in minimum potential. This charge density is well within the range of typical values for polar interfaces [S10].